\documentclass[12pt]{article}    
\usepackage[margin=1in]{geometry}
\usepackage{setspace}
\usepackage{graphicx}
\usepackage{pifont,latexsym,ifthen,theorem,rotating,calc,textcase,booktabs}
\usepackage{amsfonts,amssymb,amsbsy,amsmath}
\usepackage[errorshow]{tracefnt}
\usepackage[title]{appendix}

\usepackage{listings}
\usepackage[svgnames]{xcolor}
\usepackage{graphicx,amsmath,float,amssymb}
\usepackage{latexsym,natbib,url,hyperref}
\usepackage{wrapfig,lipsum,booktabs}
\usepackage{caption,subcaption}
\usepackage{multirow,enumitem}

\usepackage{bbm}
\usepackage{lineno}

\usepackage{epstopdf} 
\usepackage{algpseudocode}

\usepackage{gensymb} 
\usepackage{eqnarray}

\begin{document}
\title{Joint Temperature-Precipitation Patterns in the U.S. South Central Region: Multivariate Functional Inference and Gaussian Mixture Modeling}


 \author{Gihani V. W. W. Disanayakage\thanks{Texas Tech University, Lubbock, TX, USA. Email: gwickram@ttu.edu}
\and
 Asim K. Dey\thanks{Texas Tech University, Lubbock, TX, USA. Email: a.dey@ttu.edu}
\and
Kumer P. Das\thanks{University of Louisiana at Lafayette, Lafayette, LA, USA. Email:kumer.das@louisiana.edu}}


\date{}
\maketitle

\begin{abstract}
\noindent Joint variability in temperature and precipitation is central in characterizing seasonal climate structure and associated environmental processes, yet many regional analyses rely on marginal or univariate summaries. We analyze seasonal temperature–precipitation patterns across the Southern United States using two complementary multivariate statistical approaches. First, functional multivariate analysis of variance (FMANOVA) is employed to test the equality of state‑level bivariate mean functions, with the permutation‑based Wilks' lambda and Pillai's trace statistics. Second, Gaussian mixture models are applied to station‑level seasonal summaries to identify latent climate regimes based on the joint distribution of temperature and precipitation. The FMANOVA results indicate statistically significant differences in bivariate mean trajectories between states in both winter and summer, with seasonal contrasts reflecting differing contributions of temperature and precipitation. Clustering analysis indicates more clearly defined and spatially coherent winter regimes than summer regimes, with summer regimes exhibiting greater variability and a stronger role for precipitation.
\end{abstract}

\textbf{Key Words:} Joint temperature and precipitation dynamics; Functional multivariate analysis of variance; Clustering; Gaussian mixture model; Spatiotemporal climate clusters.

\maketitle

\section{Introduction}

Climate change is one of the most pressing challenges facing humanity, threatening the stability of the social and environmental systems that sustain both natural ecosystems and human communities. Understanding how key climate variables, particularly temperature and precipitation, co-vary under global warming is crucial because their joint behavior governs hydro-meteorological hazards and downstream societal impacts, including flooding, heavy-rainfall-triggered landslides, soil erosion, and runoff-driven urban impacts~\citep{ECK2020108053,Danielle_2022,Cardell_2020,gori2022_tc_rainfall_surge,lesk2022_compound_heat_moisture, Intergovernmental_Panel_on_Climate_Change_(IPCC)_2023, DHARMARATHNE2024102123}. 
Recent assessments conclude that human influence has unequivocally warmed the atmosphere, ocean, and land and is already altering baseline conditions and extremes~\citep{naz2019effects,Megan_2020,ipcc_ar6_wgi_spm_2021}. In the United States, the Fifth National Climate Assessment documents long-term warming and increases in heavy precipitation. In the Southern Great Plains, including areas overlapping the West and East South Central states, annual mean temperatures have risen since 1900, and the frequency of daily precipitation events exceeding 2 inches has increased across Oklahoma and Texas. In the Southeast, the assessment identifies increasing intensity of climate stressors, including extreme precipitation and persistent drought, and projects that future increases in total precipitation will be driven primarily by enhanced event-level extremes rather than changes in mean precipitation~\citep{mcpherson2023_nca5_ch26}.

A large literature models temperature and precipitation anomalies using separate univariate time series models with deterministic or stochastic trend components, providing evidence of persistence, nonstationarity, and marked heterogeneity across U.S. states and regions \citep{cli4010010, shen2016six, gil2022temperature}. Related studies employ univariate forecasting and projection frameworks for regional temperature and precipitation series, typically emphasizing long-run trend behavior and low-frequency variability \citep{lai2020use}. Subsequent methodological developments incorporate stochastic trends, autoregressive dependence, and regime-dependent structures to accommodate evolving distributional properties in U.S. temperature and precipitation records \citep[e.g.,][]{knutti2016climate, franzke2015climate}. 
In parallel, much of the hazards-oriented literature employs single variable extremes models, most commonly for precipitation, including analyses of trends in extreme rainfall frequency across the contiguous United States~\citep{armal2018trends,Miniussi_2020,Dey02012016,Tania_2020,gori2022_tc_rainfall_surge} and spatial extremes models used to estimate precipitation return levels over U.S. domains~\citep{cooley2010spatial,Martin_2021,Zhong02012025}. Temperature-only analyses are likewise widespread in heat-related research~\citep{perkins2013measurement, barriopedro2023heat, fard2023spatio}, and surveys of heat wave definitions note that most operational metrics are based
on temperature or temperature-derived indices rather than the joint temperature-precipitation structure~\citep{bunting2024heat}. 

Existing work on climate regimes and weather types has largely relied on univariate seasonal indices or has analyzed temperature and precipitation separately, which can obscure their joint seasonal evolution and dependence structure~\citep{region2017quarterly}. Clustering and related unsupervised learning approaches are widely used in climatology to define circulation types and weather regimes, providing low-dimensional descriptions of high-dimensional atmospheric variability. However, when the scientific objective is impact-relevant surface climate, two gaps are especially salient. First, regimes derived from large-scale circulation fields do not necessarily preserve the joint surface temperature–precipitation behavior that governs many climate hazards. Second, many operational summaries remain marginal, often univariate, treating temperature and precipitation independently and thereby obscuring compound conditions (e.g., warm–dry or cool–wet seasons) that arise from their co-variability. Together, these limitations motivate data-driven climate classifications that explicitly model multivariate surface features and their dependence structure, yielding interpretable regime structures.

The aim of this study is to conduct a multivariate functional climate‑regime analysis that characterizes seasonal patterns in the joint behavior of temperature and precipitation and leverages these patterns to compare and interpret regional climate variability. To this end, we employ a data‑driven strategy consisting of two main components. 
First, focusing on winter and summer 2025, we apply multivariate functional analysis of variance (FMANOVA) to daily temperature–precipitation curves to test for differences in state-level mean functions. Significant effects are further examined through pairwise state comparisons and variable-level attribution analyses to determine whether temperature, precipitation, or both variables drive each contrast.
Second, we cluster station-level bivariate seasonal means of temperature and precipitation for winter and summer over the period 2020–2025 across seven U.S. states in the West and East South Central regions (Texas, Oklahoma, Louisiana, Arkansas, Mississippi, Tennessee, and Alabama). Clustering is conducted separately for each year and season using Gaussian mixture models, with the number of components and covariance structures selected via the Bayesian Information Criterion (BIC). This approach yields a small set of interpretable climatic regimes (e.g., warm-dry and cool-wet) that effectively summarize station-level seasonal conditions.


The remainder of this paper is organized as follows. Section~\ref{Sec:Data} describes the data, and section~\ref{Sec:Methodology} describes the FMANOVA and  Gaussian mixture model based clustering methodology. Section~\ref{sec:results} presents the resulting climate comparisons, with FMANOVA results and clustering results. Finally, Section~\ref{sec:Concl} provides concluding remarks and discusses the broader implications of the findings.

\section{Data}\label{Sec:Data}

We obtain the climate data set from the \textit{National Centers for Environmental Information (NCEI)} through the Climate Data Online (CDO) portal~\citep{NOAA_NCEI_CDO_2025}. The data consists of daily observations of \textit{air temperature} and \textit{precipitation} recorded at multiple weather stations across seven states in the U.S. West South Central and East South Central divisions. These states include Texas, Oklahoma, Louisiana, Arkansas, Mississippi, Tennessee, and Alabama. We focus our analysis on the winter and summer seasons in 2025. All analyses adopted meteorological season definitions, with winter spanning from December 1 to February 28 and summer spanning from June 1 to August 31.

Although each state contains numerous weather stations, we limited our analysis to those with complete records for both variables across all days within the selected periods. Appendix A, Figure~\ref{fig:Station_locations} shows the spatial distribution of the stations included in the study, where each dot corresponds to a single station, and colors denote different states. Texas contains the largest number of stations (22), followed by Mississippi and Tennessee (6 each), Alabama (5), Oklahoma (4), and Louisiana and Arkansas (2 each). In total, 44 stations were included in the analysis. To facilitate a clear assessment of climatic variability across the southern U.S., we focus our analysis on two meteorologically distinct seasons, i.e., winter (December-February) and summer (June-August).

\section{Methodology}
\label{Sec:Methodology}

\subsection{Analysis of Variance for Multivariate Functional Data}

For each station $j$ in state $i$, we observe independent bivariate functional observations
\[
\mathbf{X}_{ij}(t) = \big( X_{ij1}(t), X_{ij2}(t) \big)', \quad t = 1, \dots, T,\ \ i = 1, \dots, \ell,\ \ j = 1, \dots, n_i,
\]
where $X_{ij1}(t)$ denotes the daily average air temperature, $X_{ij2}(t)$ denotes the daily total precipitation, and $T$ is the total number of days in the observation period. The number of states is $\ell = 7$. For each state $ i $, we define the \textit{bivariate} mean function as
    \[ 
        \boldsymbol{\mu}_{i}(t) = \mathbb{E}\!\left[\textbf{X}_{ij}(t)\right] = (\mu_{i1}(t), \mu_{i2}(t))',
    \]

where $\mu_{i1}(t) = \mathbb{E}\!\left[X_{ij1}(t)\right] $ is the state-level mean temperature on day $ t $ and $\mu_{i2}(t) = \mathbb{E}\!\left[X_{ij2}(t)\right]$ is the state-level mean precipitation on day $t$. For each day \(t\), the vector \(\boldsymbol{\mu}_{i}(t)\) is computed as the cross-station average of the temperature and precipitation observations recorded in state \(i\). We consider two seasonal intervals, \(I\), corresponding to Winter and Summer.

We assume that these random functions $\mathbf{X}_{ij}$ are independent and identically distributed within each state. Under this framework, we conduct a functional multivariate analysis of variance (FMANOVA) to test the equality of mean functions across all $\ell$ states as
\begin{equation}
H_0:\ \boldsymbol{\mu}_1(t) = \boldsymbol{\mu}_2(t) = \cdots = \boldsymbol{\mu}_\ell(t), 
\label{eq:H0}
\end{equation}
against the alternative that at least one state differs in its mean temperature–precipitation curve over the considered period.


Before applying FMANOVA, we represent the bivariate curve at each station using a finite set of orthonormal basis functions, $\{\boldsymbol{\psi}_r\}_{r=0}^{K_m}$. The components of $\textbf{X}_{ij}(t)$ are expressed as
    \begin{equation}
        \textbf{X}_{ij}(t)\ \approx\
        \begin{bmatrix}
            \boldsymbol{\alpha}_{ij1}\\[-0.2em]
            \boldsymbol{\alpha}_{ij2}
        \end{bmatrix}
    \boldsymbol{\psi}(t)\ =\ \boldsymbol{\alpha}_{ij}\,\boldsymbol{\psi}(t),\qquad t\in I.
    \label{eq:basis}
    \end{equation}
where $\boldsymbol{\alpha}_{ijm} = \big(\alpha_{ijm1},\dots,\alpha_{ijmK_m,0,\dots,0})^{\!\top}\in\mathbb{R}^{KM}$, $m=1,2$, $KM=max\{K_1,K_2\}$. We represent the curves using \textit{B-spline} basis functions. A spline function is built by piecing together low-degree polynomials on subintervals defined by a set of knots, with the pieces joined so that the overall function is smooth at the knots. The associated \textit{B-spline} functions provide a convenient local basis: for a given order and knot sequence, any spline can be written uniquely as a linear combination of these B-spline basis functions~\citep{deboor1978practical,schumaker2007spline}. For each component $m$, the coefficients in $\mathbf{\alpha}_{ij}$ are estimated using the least squares method. The optimal value of $K_m$ is selected using the Bayesian Information Criterion (BIC)~\citep{shmueli2010explain}. We then take the modal value of $K_m$ across all processes as the common choice for all curves $(X_{ijm}(t),\, t \in I),\ i = 1,\dots,l,\ j = 1,\dots,n_i$.

We then introduce the following matrices, which will be used to construct the test statistics for FMANOVA for Eq.~\ref{eq:H0}. Let $ \mathbf{\bar{X}}_{i}(t) = (1/n_i) \sum_{j=1}^{n_i}\mathbf{X}_{ij}(t), i = 1, \dots, l $ and $ \mathbf{\bar{X}}(t) = (1/n) \sum_{j=1}^{n_i} \mathbf{X}_{ij} (t) $, $ t \in I $. We define

    \[
        \textbf{E} = \sum_{i=1}^{\ell}\sum_{j=1}^{n_i}\int_I\!\big(X_{ij}(t)-\bar X_i(t)\big)\big(X_{ij}(t)-\bar X_i(t)\big)^{\!\top}\,dt,
    \]
    
    \[
        \textbf{H} = \sum_{i=1}^{\ell}\int_I\!\big(\bar X_i(t)-\bar X(t)\big)\big(\bar X_i(t)-\bar X(t)\big)^{\!\top}\,dt.
    \]

In this study, we consider two classical MANOVA tests: \textit{Wilks’ lambda}
and \textit{Pillai’s trace}. The Wilks’ lambda statistic is defined as
\begin{equation}
    W = \frac{\det(\mathbf{E})}{\det(\mathbf{E} + \mathbf{H})},
    \label{Eq:Wilks}
\end{equation}
where \( \mathbf{E} \) and \( \mathbf{H} \) denote the error and hypothesis
sum-of-squares-and-cross-products matrices, respectively. Smaller values of \( W \)
indicate stronger evidence against the null hypothesis, as they correspond to a larger
proportion of variability explained by the hypothesis matrix \( \mathbf{H} \)~\citep{Wilks1932, Anderson1958, TODOROV201037}. Similarly, the Pillai’s trace statistic is defined as
\begin{equation}
    P = \operatorname{tr}\!\left\{ \mathbf{H} (\mathbf{H} + \mathbf{E})^{-1} \right\}.
    \label{Eq:Pillai}
\end{equation}
Pillai’s trace is often regarded as more robust
to departures from model assumptions, such as non-normality or heterogeneity of
covariance matrices~\citep{Pillai1955}. The \( W \) and \( P \) statistics are not analytically tractable in the functional data
setting; therefore, \( p \)-values are computed using permutation-based methods
\citep{Gorecki10092017}. We reject the null hypothesis \( (H_0) \) when the permutation
\( p \)-value is less than or equal to \( \alpha \), where \( \alpha \) denotes the
significance level of the test.

When reporting pairwise FMANOVA comparisons across all state pairs, we control the
false discovery rate using the Benjamini--Hochberg (BH) procedure~\citep{benjamini1995controlling}, applied to the permutation \( p \)-values. The BH procedure accounts for multiple testing by ordering the \( p \)-values and selecting a rejection threshold that controls the expected proportion of false discoveries among the rejected hypotheses at a prespecified level.


\subsection{Clustering Climate Patterns Using a Gaussian Mixture Model}
\label{sec:Clustering}

To identify dominant seasonal climate regimes, we perform model-based clustering of station-level climate summaries using Gaussian mixture models (GMM). The analysis considers $\ell=7$ states indexed by $i \in \{1,\ldots,\ell\}$, with weather stations indexed by $j = 1,\ldots,n_i$ within each state. We focus on the study year 2025 and two climatologically distinct seasons, $q \in \{\text{Winter}, \text{Summer}\}$, consistent with the notation used in the FMANOVA framework.

For each season $q$, let $I_q$ denote the set of calendar days belonging to that season in 2025, with length $L_q = |I_q|$. 
For each state $i$ and weather station $j$, we observe daily mean air temperature $X_{ij1}(t)$ and daily precipitation $X_{ij2}(t)$ for each day $t \in I_q$. To summarize seasonal climate conditions at each station, we compute bivariate seasonal means of temperature and precipitation. Specifically, for station $(i,j)$ in season $q$, we define
\[
\mathbf{x}_{ij}^{(q)} =
\begin{bmatrix}
\bar X_{ij1}^{(q)} \\
\bar X_{ij2}^{(q)}
\end{bmatrix}
=
\begin{bmatrix}
\frac{1}{L_q} \sum_{t \in I_q} X_{ij1}(t) \\
\frac{1}{L_q} \sum_{t \in I_q} X_{ij2}(t)
\end{bmatrix}.
\]
For a given season $q$, we denote by $\mathbf{x}^{(q)} \in \mathbb{R}^2$ the collection $\{\mathbf{x}_{ij}^{(q)}\}$ across all states and stations. This collection represents the full station-level dataset of seasonal climate features used for clustering. We model the distribution of $\mathbf{x}^{(q)}$ using a $K$-component Gaussian mixture model of the form
\begin{equation}
p\!\left(\mathbf{x}^{(q)}\right)
=
\sum_{k=1}^K \pi_k \,
\mathcal{N}\!\left(\mathbf{x}^{(q)} \mid \boldsymbol{\mu}_k, \boldsymbol{\Sigma}_k \right),
\label{Eq:GMM}
\end{equation}
where $\pi_k > 0$ are the mixing proportions satisfying $\sum_{k=1}^K \pi_k = 1$, $\boldsymbol{\mu}_k$ is the mean vector of component $k$, and $\boldsymbol{\Sigma}_k$ is its covariance matrix. Each component represents a latent climate regime characterized by a distinct joint distribution of seasonal mean temperature and precipitation.

Component membership is represented through a latent indicator vector $\mathbf{z} = (z_1,\ldots,z_K)$ with one-of-$K$ encoding, where $z_{ik} = 1$ if an observation $\mathbf{x}^{(q)}$ arises from component $k$ and $z_k = 0$ otherwise. The prior distribution of $\mathbf{z}$ is governed by the mixing proportions, $p(\mathbf{z}) = \prod_{k=1}^K \pi_k^{z_k}$, and the conditional distribution of $\mathbf{x}^{(q)}$ given $\mathbf{z}$ is
\[
p(\mathbf{x}^{(q)} \mid \mathbf{z})
=
\prod_{k=1}^K
\mathcal{N}\!\left(\mathbf{x}^{(q)} \mid \boldsymbol{\mu}_k, \boldsymbol{\Sigma}_k \right)^{z_k}.
\]
The posterior probability that an observation belongs to component $k$, referred to as the \emph{responsibility}, is obtained via Bayes’ theorem as
\begin{equation}
\gamma(z_k)
=
\Pr(z_k = 1 \mid \mathbf{x}^{(q)})
=
\frac{\pi_k \, \mathcal{N}\!\left(\mathbf{x}^{(q)} \mid \boldsymbol{\mu}_k, \boldsymbol{\Sigma}_k \right)}
{\sum_{h=1}^K \pi_h \, \mathcal{N}\!\left(\mathbf{x}^{(q)} \mid \boldsymbol{\mu}_h, \boldsymbol{\Sigma}_h \right)}.
\label{Eq:responsibilities}
\end{equation}

Each component covariance matrix $\boldsymbol{\Sigma}_k$ is parameterized using an eigen-decomposition, $\boldsymbol{\Sigma}_k = \lambda_k \mathbf{D}_k \mathbf{A}_k \mathbf{D}_k',$
where $\mathbf{D}_k$ determines the orientation of the cluster, $\mathbf{A}_k$ is a diagonal matrix governing its shape, and $\lambda_k > 0$ controls its volume. This flexible parameterization allows clusters to differ in size, shape, and orientation, which is well-suited for capturing heterogeneous climate regimes. We can estimate model parameters $\Theta = \{\pi_k, \boldsymbol{\mu}_k, \boldsymbol{\Sigma}_k\}_{k=1}^K$ by maximum likelihood using the Expectation--Maximization (EM) algorithm~\citep{Dempster_2018, Meng_2002}. The observed-data log-likelihood is
\[
l(\Theta)
=
\sum_{r=1}^{n}
\log\!\left(
\sum_{k=1}^K
\pi_k \,
\mathcal{N}\!\left(\mathbf{x}_r^{(q)} \mid \boldsymbol{\mu}_k, \boldsymbol{\Sigma}_k \right)
\right),
\]
where $n = \sum_{i=1}^\ell n_i$ is the total number of stations. The EM algorithm alternates between an E-step and an M-step, which updates the parameters as follows:
\begin{itemize}
    \item \textbf{\emph{E-step:}} Computes the responsibilities $ \gamma_{k} $ defined in equation~\ref{Eq:responsibilities} which represent the probability that observation $\mathbf{x}^{(q)}$ belongs to latent Gaussian component $k$.
    \item \textbf{\emph{M-step:}} Re-estimate the parameters using the current $\gamma_{k}$ values by: 
    \[
\widehat{\boldsymbol{\mu}}_k =
\frac{1}{\widehat{N}_k}
\sum_{r=1}^n \gamma_{rk} \mathbf{x}_r^{(q)},
\qquad
\widehat{\boldsymbol{\Sigma}}_k =
\frac{1}{\widehat{N}_k}
\sum_{r=1}^n \gamma_{rk}
(\mathbf{x}_r^{(q)} - \widehat{\boldsymbol{\mu}}_k)
(\mathbf{x}_r^{(q)} - \widehat{\boldsymbol{\mu}}_k)',
\]
with $\widehat{N}_k = \sum_{r=1}^n \gamma_{rk}$ and $\widehat{\pi}_k = \widehat{N}_k / n$. 
\end{itemize}

Iteration continues until convergence of the log-likelihood~\citep{bishop2006prml}. We perform GMM clustering using the \textit{mclust} R-package~\citep{mclust}. For each candidate $ K \in \{1, \dots, K_{max} \} $, we run the EM algorithm and retain the solution with the highest observed-data log-likelihood. The optimal number of components $K$ is selected by the Bayesian Information Criterion (BIC), $\mathrm{BIC}= 2\,l(\widehat\Theta) - m \log n,$ where $l(\widehat\Theta)$ is the maximized log-likelihood, $m$ is the number of estimated parameters, and $ n $ is the sample size. In general, the larger the value of the BIC, the stronger evidence for the model and number of clusters~\citep{Fraley01062002}.

\section{Results}
\label{sec:results}

\subsection{Climate comparison}

\begin{figure*}[!ht]
     \centering
         \begin{subfigure}[b]{0.49\textwidth}
    \centering
      \includegraphics[width=1.0\textwidth]{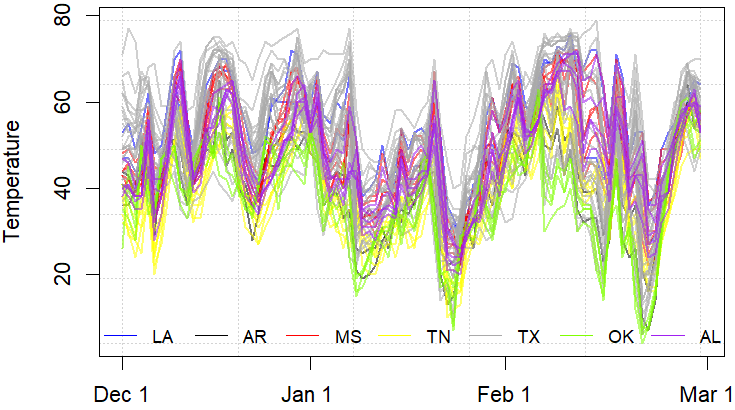}
    \caption{Winter 2025.}
     \end{subfigure}
   \begin{subfigure}[b]{0.49\textwidth}
    \centering
    \includegraphics[width=1.0\textwidth]{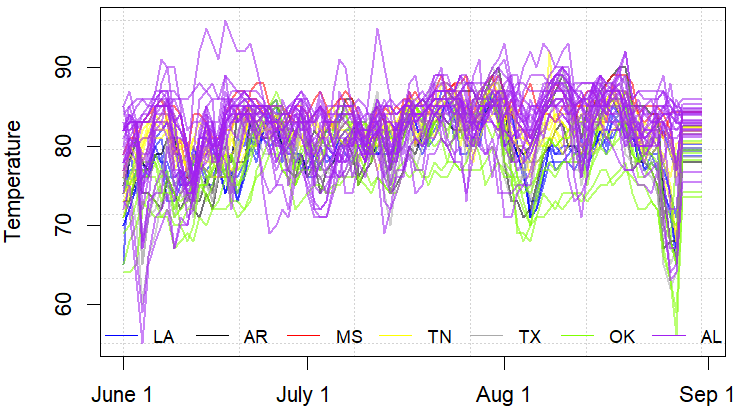}
    \caption{Summer 2025.}
     \end{subfigure}  
    \begin{subfigure}[b]{0.49\textwidth}
    \centering
      \includegraphics[width=1.0\textwidth]{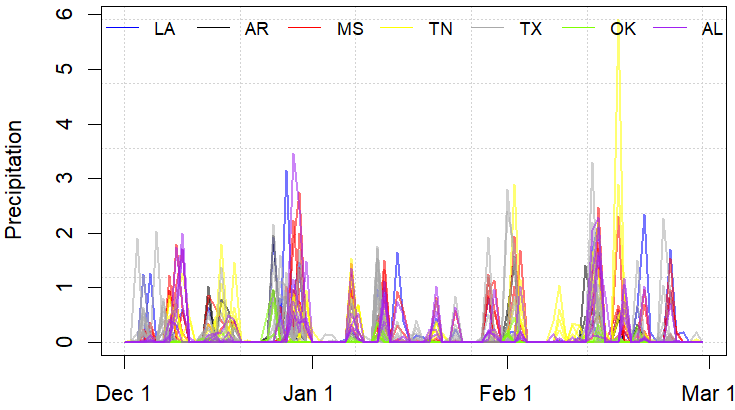}
    \caption{Winter 2025.}
     \end{subfigure}
   \begin{subfigure}[b]{0.49\textwidth}
    \centering
    \includegraphics[width=1.0\textwidth]{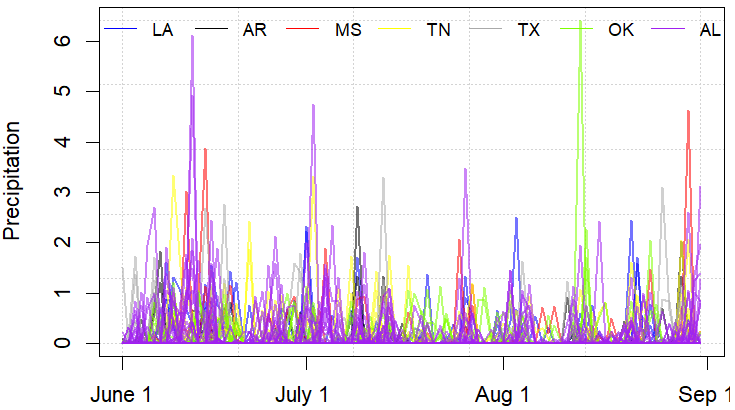}
    \caption{Summer 2025.}
     \end{subfigure}  
     \caption{Daily average temperature (top row) and daily total precipitation (bottom row) across seven states in 2025.}    
    \label{fig:PCP_Temp_Plots}
\end{figure*}

The top row of Figure~\ref{fig:PCP_Temp_Plots} displays daily mean air temperatures from all selected stations across TX, OK, LA, AR, MS, TN, and AL. The left and right panels correspond to the winter and summer of 2025, respectively. Each curve represents one station, colored by state. Vertical spread on a given day reflects spatial variability, while within-curve fluctuations indicate day-to-day changes. Winter temperatures range from roughly 10\degree F to 80\degree F and show pronounced spatial and temporal variability, with colder conditions in northern and inland states and warmer temperatures along the Gulf Coast. Summer temperatures cluster tightly between 70\degree F and 90\degree F, with reduced inter-station spread and more homogeneous behavior across the region.

The bottom row of Figure~\ref{fig:PCP_Temp_Plots} shows daily total precipitation across stations for winter (left) and summer (right) of 2025. Most days exhibit near-zero precipitation, with intermittent sharp peaks corresponding to precipitation events. In winter, these peaks are infrequent but often intense and synchronized across stations, reflecting region-wide frontal systems. Summer displays more frequent nonzero totals and dense layering near the baseline, with scattered short spikes arising from localized convective storms. Overall, winter precipitation appears episodic and spatially coherent, whereas summer features more frequent, localized events superimposed on a generally wetter background.
Taken together, Figures~\ref{fig:PCP_Temp_Plots} highlight a clear seasonal contrast: winter shows substantial temperature variability and intermittent synoptic-scale precipitation, while summer is marked by consistently warm conditions and localized rainfall. This motivates separate modeling of winter and summer in the subsequent functional analyses.

\subsection{Functional MANOVA outputs}

In this section, we analyze bivariate functional data from the selected weather stations in 7 states for winter and summer 2025. For each season $\{\text{Winter},\text{Summer}\}$, we separately apply FMANOVA to test the equality of the bivariate mean functions (temperature and precipitation) across states. We then identify which pairs of states exhibit significant differences and determine whether these differences are driven primarily by temperature, precipitation, or both variables.

Table~\ref{table:FMANOVA_Test_statistics} presents the FMANOVA results for Wilk's test and Pillai test. We find that for both winter and summer, the $p$-values of the Wilks' test and Pillai test are less than 0.01. Therefore, we reject the null hypothesis and conclude that the 7 states differ significantly in their bivariate mean temperature-precipitation curves in both winter and summer in 2025.

\begin{table*}[!ht]
\caption{FMANOVA test statistics results.}
\label{table:FMANOVA_Test_statistics}	
\centering
\begin{tabular}{l|lcccc}  
    \hline		
{Season} & {Test} &   {Test Statistics}  &     $p$-value\\
        \hline
	Winter    & Wilks test & 0.347 & $<$ 0.001 \\
        & Pillai test & 0.814 & $<$ 0.001  \\
        \hline
       Summer & Wilks test(W) & 0.401 & $<$ 0.001 \\
        & Pillai test(P) & 0.709 & $<$ 0.001 \\
        \hline
	\end{tabular}
\end{table*}

We implement the FMANOVA tests using the \texttt{fdANOVA} R package~\citep{Gorecki_2019}, which provides permutation-based procedures for comparing mean functions of functional data. Functional observations are represented using a cubic B-spline basis (order 4), and permutation $p$-values are computed using 20000 permutations.


We subsequently conduct pairwise FMANOVA comparisons for all state pairs and obtain BH–adjusted $p$-values for each test. Table~\ref{table:Pairwise_comparison} reports the adjusted $p$-values and identifies the state pairs that drive the rejection of the null hypothesis in Eq.~\ref{eq:H0} at the $\alpha=0.05$ significance level. In winter, statistically significant differences are mainly observed between geographically distant states, such as TN--TX, MS--TX, and OK--TN, whereas comparisons among neighboring or climatically similar states (e.g., AL--AR, AR--LA, and LA--MS) are generally not
significant. This pattern suggests a broad spatial gradient, with stronger contrasts between northern or inland states and southern or Gulf-adjacent states. In summer, a larger number of pairwise differences emerge, including several involving Gulf-coast states (e.g., AL--LA, AL--MS, LA--OK, MS--TN, and TN--TX), indicating more localized yet spatially structured heterogeneity. Overall, the results point to a clear spatial organization in temperature dynamics, with winter differences driven by large-scale regional contrasts and summer differences reflecting finer-scale spatial variation across the study region.

\begin{table*}[ht]
\caption{Pairwise Comparison of States.}
\label{table:Pairwise_comparison}
\centering
\begin{tabular}{lcc}
\toprule
\multirow{2}{*}{Contrast} & \multicolumn{2}{c}{$p$-value (BH–adjusted)} \\ 
\cmidrule(lr){2-3}
 & Winter & Summer \\
\midrule
	    AL -- AR & 0.879 & 0.070\\
        AL -- LA & 0.879 & 0.048\\
        AL -- MS &  0.429 & 0.015\\
	    AL -- OK & 0.825 & 0.015\\
        AL -- TN & 0.012 & 0.177\\
        AL -- TX & 0.065 & 0.019\\
        AR -- LA & 0.289 & 0.385\\
        AR -- MS & 0.380 & 0.064\\
        AR -- OK & 0.748 & 0.889\\
        AR -- TN & 0.068 & 0.158\\
        AR -- TX & 0.282 & 0.385\\
        LA -- MS & 0.740 & 0.019\\
        LA -- OK & 0.713 & 0.048\\
        LA -- TN & 0.068 & 0.048\\
        LA -- TX & 0.068 & 0.185\\
        MS -- OK & 0.294 & 0.015\\
        MS -- TN & 0.016 & 0.008\\
        MS -- TX & 0.009 & 0.385\\
        OK -- TN & 0.012 & 0.064\\
        OK -- TX & 0.219 & 0.043\\
        TN -- TX & 0.001 & 0.007\\
        \hline
	\end{tabular}
\end{table*}

Table~\ref{tab:Most_influential_variable_selection} in Appendix A breaks down the significant multivariate state–state differences reported in Table~\ref{table:Pairwise_comparison} into their temperature and precipitation components. For each significant pair, we assess temperature (TAVG) and precipitation (PRCP) separately and report the BH‑adjusted \textit{p}-values. This decomposition clarifies whether the detected differences arise from temperature, precipitation, or a combination of both. 
In winter, three pairs (AL-TN, MS-TN, MS-TX) show BH-adjusted \textit{p-values} below $0.05$ for both variables, indicating that differences in both temperature and precipitation contribute to the multivariate separation. The remaining two pairs (OK-TN and TN-TX) are driven mainly by temperature, with non-significant precipitation effects. In summer, most significant pairs are primarily temperature-driven, with only four pairs (AL-OK, MS-OK, OK-TX, TN-TX) showing contributions from both variables; no pair shows a precipitation-only effect.


\subsection{GMM Clustering Results}

\subsubsection{Clustering Climate Regimes in 2025}

In this section, we use Gaussian mixture modeling to cluster the bivariate means for each station and season in 2025. For winter and summer separately, we fit GMMs with varying numbers of components and select the optimal number of clusters and the structure using the model-selection criteria described in Section~\ref{sec:Clustering}. Each station is then assigned to its most likely cluster, yielding distinct seasonal climate regimes.

\begin{figure*}[!ht]
     \centering
     \begin{subfigure}[b]{0.49\textwidth}
         \centering
       \includegraphics[width=0.85\textwidth]{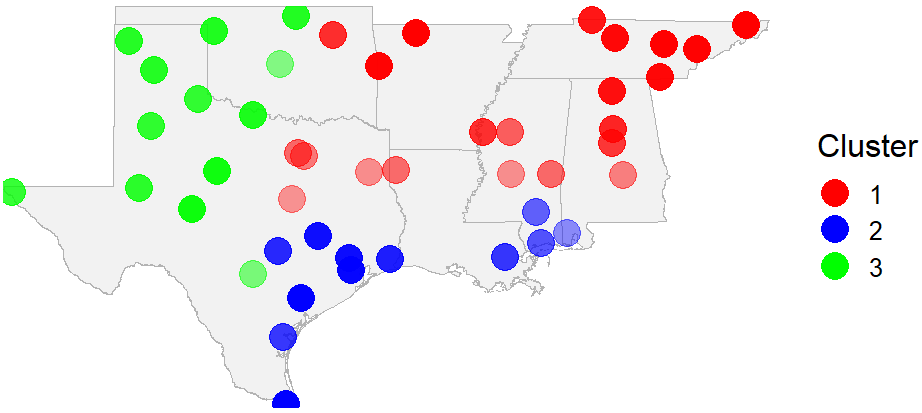}
       \caption{Winter 2025.}
     \end{subfigure}
       \begin{subfigure}[b]{0.49\textwidth}
         \centering
       \includegraphics[width=0.85\textwidth]{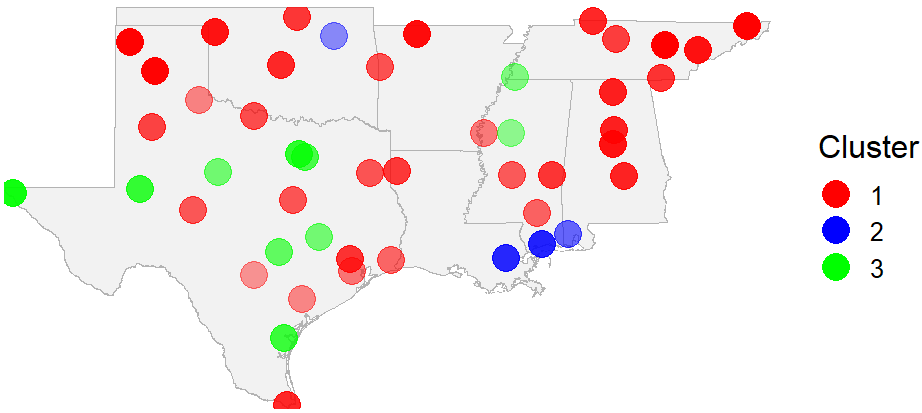}
       \caption{Summer 2025.}
     \end{subfigure}
\caption{Gaussian Mixture Model clustering of station–summer climate features.}
\label{fig:clusters1}
\end{figure*}

Figure~\ref{fig:clusters1} shows the results of Gaussian mixture model clustering applied to station-level temperature and precipitation for winter and summer 2025 across Texas, Oklahoma, Arkansas, Louisiana, Mississippi, Tennessee, and Alabama. Each point represents a weather station and is colored according to its assigned cluster, with panels (a) and (b) corresponding to winter and summer, respectively. In winter, the clustering exhibits clear spatial structure, with stations in northern and inland regions forming clusters distinct from those in southern and Gulf-adjacent areas. The strong geographic coherence of the clusters suggests that winter climate variability across the study region is driven by large-scale synoptic processes and pronounced temperature and precipitation gradients.

In summer, the clustering pattern is less spatially segregated, with greater overlap of clusters across states. Although some regional differentiation remains, particularly between inland and coastal stations, the separation among clusters is weaker than in winter. This reflects the more homogeneous summer climate conditions across the region, with reduced large-scale gradients and greater influence of localized, short-lived weather processes. 


\begin{figure*}[!ht]
     \centering
     \begin{subfigure}[b]{0.49\textwidth}
         \centering
       \includegraphics[width=1.0\textwidth]{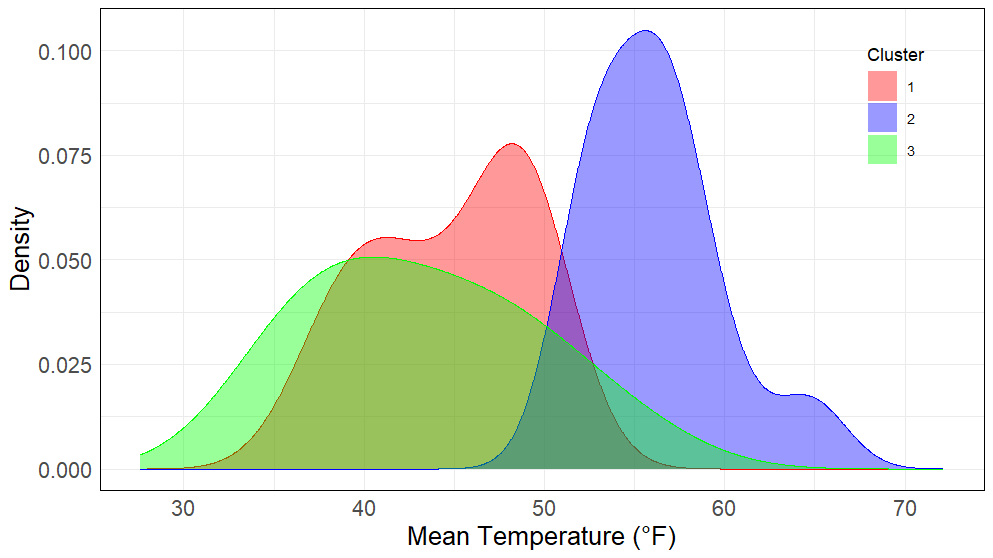}
       \caption{Temperature, Winter 2025}
     \end{subfigure}  
     \begin{subfigure}[b]{0.49\textwidth}
         \centering
       \includegraphics[width=1.0\textwidth]{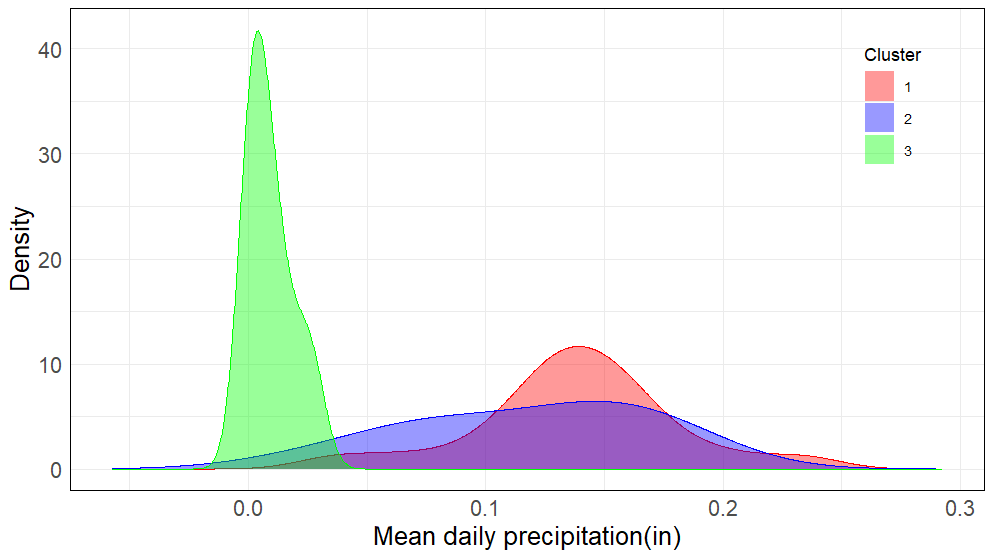}
       \caption{Precipitation, Winter 2025.}
     \end{subfigure}
          \begin{subfigure}[b]{0.49\textwidth}
         \centering
       \includegraphics[width=1.0\textwidth]{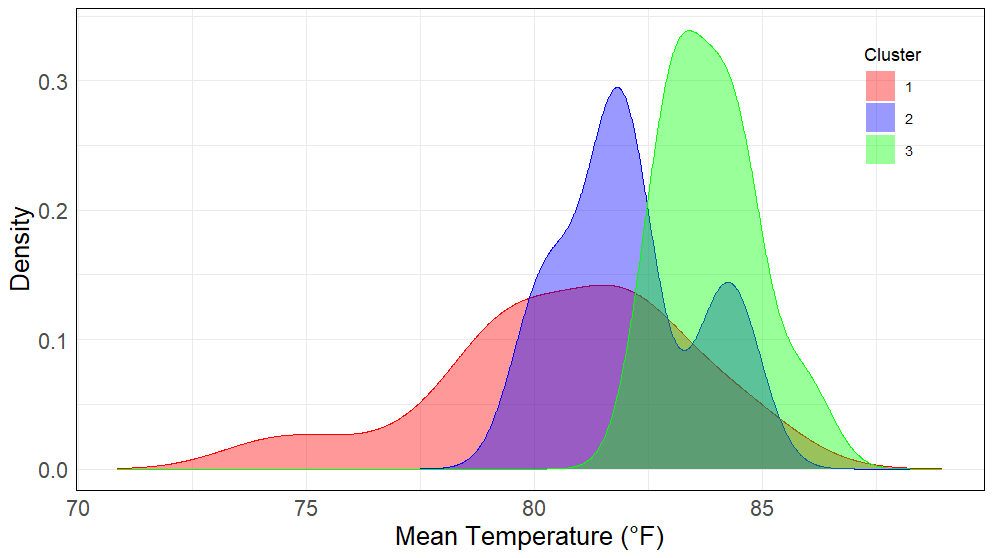}
       \caption{Temperature, Summer 2025}
     \end{subfigure}
       \begin{subfigure}[b]{0.49\textwidth}
         \centering
       \includegraphics[width=1.0\textwidth]{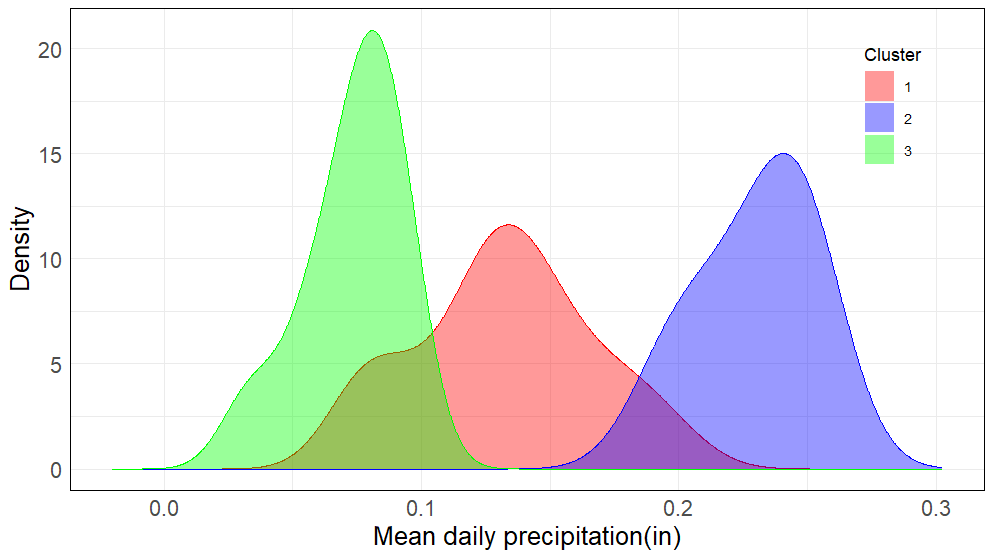}
       \caption{Precipitation, Summer 2025.}
     \end{subfigure}
\caption {Univariate density plots.}
\label{fig:univaryate_density_plots_2025}
\end{figure*}

Figure~\ref{fig:univaryate_density_plots_2025} displays the univariate density estimates of temperature and precipitation for the three clusters identified by the Gaussian mixture model, shown separately for winter and summer 2025. In winter, the temperature distributions (panel~a) exhibit clear separation across clusters, with Cluster~3 centered at lower temperatures, Cluster~1 capturing intermediate conditions, and Cluster~2 characterized by higher mean temperatures. This ordering reflects systematic climatic differences among stations and is consistent with the spatial organization observed in the clustering results. The corresponding winter precipitation densities (panel~b) also reveal distinct cluster-specific behavior, with Cluster~3 concentrated near low precipitation levels and others exhibiting progressively higher means and broader variability, indicating heterogeneity in winter precipitation regimes.

In summer, the temperature distributions (panel~c) shift uniformly toward higher values, with reduced overlap among clusters relative to winter, suggesting more stable and homogeneous warm-season conditions. Nevertheless, cluster-specific differences remain evident, with Cluster~3 consistently warmer and Cluster~1 slightly cooler on average. The summer precipitation densities (panel~d) show stronger separation than temperature, with clusters distinguished primarily by differences in both the magnitude and spread of the precipitation. Cluster~3 exhibits low and tightly concentrated precipitation, while Cluster~1 and~2 display higher means and increased variability. 


\begin{figure*}[!ht]
     \centering
     \begin{subfigure}[b]{0.49\textwidth}
         \centering
       \includegraphics[width=1.0\textwidth]{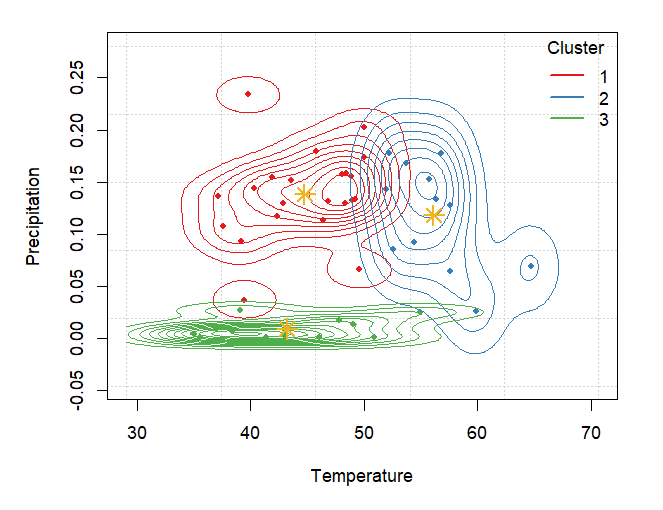}
       \caption{Winter 2025.}
     \end{subfigure}
       \begin{subfigure}[b]{0.49\textwidth}
         \centering
       \includegraphics[width=1.0\textwidth]{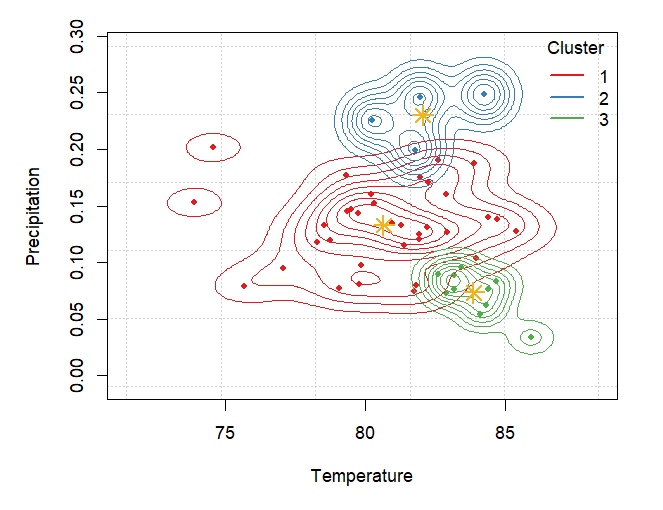}
       \caption{Summer 2025.}
     \end{subfigure}
\caption {Bivaraite contour plots: distribution of climate variables in different clusters.}
\label{fig:contour_plots_2}
\end{figure*}

Figure~\ref{fig:contour_plots_2} displays bivariate kernel density contour plots of temperature and precipitation for the three Gaussian mixture model clusters in winter and summer 2025. In winter (panel~a), the clusters exhibit clear separation in the joint temperature–precipitation space. Cluster~3 is concentrated at lower temperatures and low precipitation, reflecting colder and drier inland conditions. Cluster~1 occupies an intermediate regime with moderate temperatures and higher precipitation variability, while Cluster~2 is characterized by warmer temperatures and a broader spread in precipitation. The limited overlap among the high-density regions indicates that the clusters capture distinct winter climate regimes, with temperature playing a primary role in separation and precipitation contributing additional structure. In summer (panel~b), the joint distributions shift toward higher temperatures, and the overall separation among clusters is reduced compared with winter. Cluster~3 remains associated with relatively high temperatures and lower precipitation, whereas Cluster~2 exhibits higher precipitation levels and increased variability, consistent with humid, Gulf-influenced conditions. Cluster~1 occupies an intermediate position in both dimensions, with moderate temperatures and precipitation. Although some overlap is evident across clusters, particularly in temperature, the contour plots indicate that precipitation contributes more strongly to cluster differentiation during summer. 

\subsubsection{Spatiotemporal Dynamics of Climate Clusters}

In this section, we extend the GMM clustering analysis to examine the temporal evolution of the winter and summer climate regimes across the study region from 2020 to 2025. Consistent with the 2025 analysis, we retrieve multi-year daily climate data for 2020–2024 from the NCEI Climate Data Online (CDO) portal~\citep{NOAA_NCEI_CDO_2025}. Throughout, we use the meteorological definitions of the seasons (winter: December–February; summer: June–August). By comparing year-specific clustering patterns, we assess the stability and interannual variability of climate structures derived from temperature–precipitation patterns over time.

Figure~\ref{fig:winter_clustering2020_2025} in Appendix~B shows the GMM clusters for the winter season in the study region from 2020 to 2025. We find that the results of the winter clustering reveal a high degree of temporal persistence in the spatial organization of climate regimes over the six years. Across years, cluster membership remains broadly stable, with Cluster~2 consistently dominating the southern portion of the domain, particularly across coastal Texas, Louisiana, Mississippi, and Alabama, reflecting warmer and wetter winter conditions. Cluster~1 is repeatedly concentrated in the northern and inland areas, including Tennessee, Arkansas, and parts of Oklahoma, corresponding to cooler and relatively drier winter climates. Cluster~3 appears more localized and episodic, often occupying transitional zones between the other two clusters, suggesting sensitivity to interannual variability. While minor year-to-year shifts in station-level assignments are evident, particularly along cluster boundaries, the overall spatial coherence of winter clusters indicates that large-scale synoptic and regional climatic controls dominate interannual variability during the cold season.

In contrast, the summer clustering patterns (Figure~\ref{fig:summer_clustering2020_2025} in Appendix~B) exhibit greater temporal variability and more pronounced reorganization over years, consistent with the influence of convective precipitation, mesoscale processes, and localized extremes. Cluster~1 is frequently widespread across much of the region, especially in interior states, reflecting persistently high temperatures with moderate precipitation. Cluster~3 tends to emerge prominently in Texas and Oklahoma in several years, capturing hotter and drier summer conditions, although its spatial footprint fluctuates substantially over time. Cluster~2, often associated with higher precipitation regimes, shows less spatial stability and appears intermittently throughout the eastern portion of the domain, including Mississippi, Alabama, and Tennessee. The increased temporal fluidity of the summer clusters highlights the season’s greater sensitivity to interannual climate variability and localized forcing, underscoring the importance of joint temperature–precipitation modeling when characterizing evolving summer climate regimes in the Environmetrics context. 

Figures ~\ref{fig:smoothfun}, ~\ref{fig:cluster11}, and ~\ref{fig:cluster12} in Appendix~B show bivariate kernel density contour plots of temperature and precipitation for clusters during winter and summer in 2020, 2021, 2022, 2023, and 2024. Throughout the years, winter clusters separate primarily along the temperature axis, whereas summer clusters overlap strongly in temperature and diverge mainly through the vertical extent and right tail of precipitation. These figures highlight a consistent seasonal transition from temperature-dominated structure in winter to precipitation-driven variability in summer.

\section{Discussion}\label{sec:Concl}

This study used a multivariate functional framework to characterize recent seasonal climate variability across the West and East South Central United States, with explicit attention to the joint evolution of temperature and precipitation. By combining functional MANOVA with model‑based clustering, we provided both formal inferential comparisons of state‑level mean trajectories and a complementary data‑driven description of station‑level climate regimes. Together, these analyses demonstrate the value of multivariate, function‑aware statistical approaches for capturing compound climate structure and seasonal contrasts that are not readily apparent from marginal or univariate summaries.

The FMANOVA results show strong evidence of differences among state‑level mean temperature–precipitation curves in both winter and summer 2025 under permutation‑based Wilks' lambda and Pillai's trace tests. Pairwise comparisons reveal clear seasonal asymmetry: winter differences are fewer and are largely driven jointly by temperature and precipitation, while summer differences involve more state pairs but are more often temperature‑dominated, with precipitation contributing less frequently. The Gaussian mixture model clustering provides a complementary descriptive view of seasonal organization. In winter 2025, clusters are well separated and spatially coherent, distinguishing colder, drier inland stations from warmer, wetter Gulf‑adjacent locations, with an intermediate transitional regime. Separation is strong in univariate and joint representations, indicating a well‑defined joint temperature–precipitation structure. In summer, clusters are less distinct and less spatially organized, with greater overlap in temperature and clearer differentiation in precipitation. Extending the analysis across 2020–2025 reveals substantial persistence of winter regimes and markedly greater interannual variability in summer, particularly in precipitation‑driven clusters, highlighting seasonal contrasts in regime stability that are not captured by marginal analyses alone.

Several limitations frame directions for future work. First, the analysis focuses on a relatively short and recent time window, which limits inference on long‑term climate change and excludes formal attribution of observed patterns to secular trends rather than interannual variability. Second, clustering is performed using seasonal mean temperature and precipitation, which improves interpretability but necessarily masks intra-seasonal structure, extremes, persistence, and event‑scale dynamics that are often most relevant for impacts. Third, the FMANOVA framework assumes the independence of stations within states and does not explicitly account for spatial dependence, which may lead to understated uncertainty in states with dense station networks. Finally, the focus on two climate variables, while central, omits related processes, such as humidity, soil moisture, or large‑scale circulation indices, that may contribute to joint climate regimes.

These limitations motivate several natural statistical extensions. One direction is to expand the FMANOVA framework to longer climate records and embed explicit trend or change‑point components within the functional basis representation, enabling joint inference on the evolving mean structure and temporal nonstationarity. A second extension is to replace seasonal‑mean clustering with model‑based clustering of multivariate functional trajectories, for example, Gaussian mixture models on functional principal component scores, which would preserve intraseasonal dependence, extremes, and persistence. A third direction is to incorporate spatial dependence through hierarchical or spatially correlated functional models, such as region-level random effects or spatial covariance structures in coefficient space, which yield more realistic uncertainty quantification and improved inference under spatially dense sampling.


\section*{Declarations}
\begin{itemize}
\item Conflict of interest: The authors declare that they have no conflict of interest.
\end{itemize}

\clearpage
\begin{appendices}
\section{Data description and Climate Comparison}\label{secA01}

\begin{figure*}[!ht]
    \centering
    \includegraphics[width=0.65\linewidth]{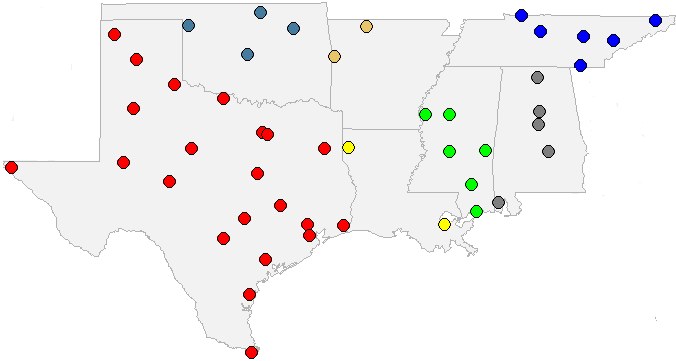}
    \caption{Station Locations.}
    \label{fig:Station_locations}
\end{figure*}

Figure~\ref{fig:Station_locations} illustrates the spatial distribution of all stations included in the study. Each dot represents an individual station, and the colors indicate the corresponding states.

\begin{figure*}[!ht]
\centering
\includegraphics[width = 0.48\textwidth]{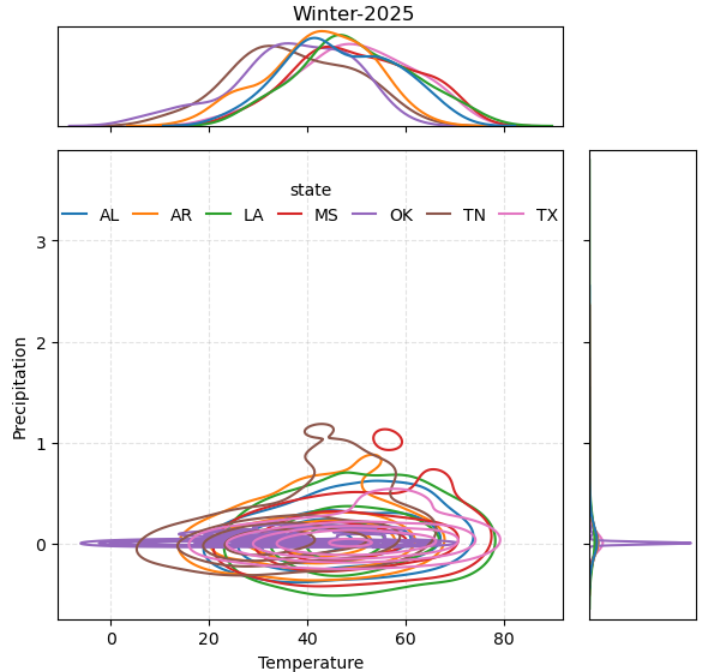}
    \includegraphics[width = 0.48\textwidth]{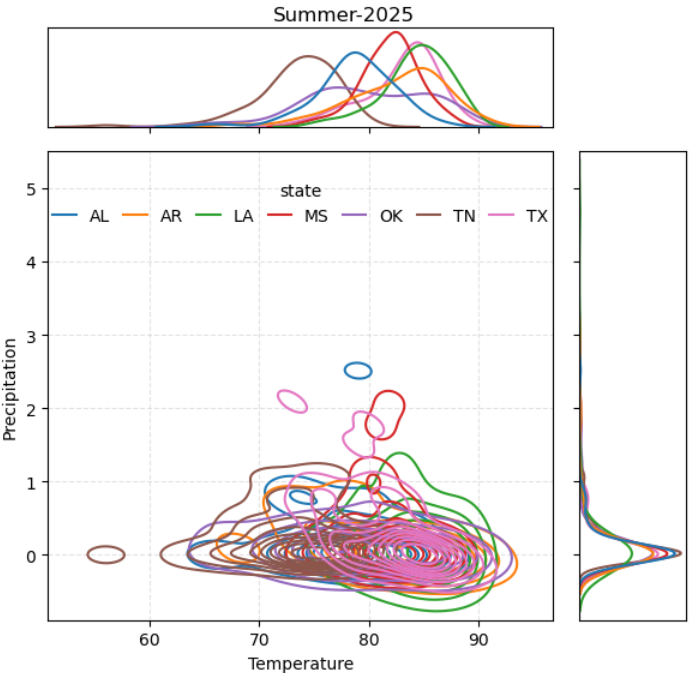}
\caption {Bivariate Density Contours of Temperature-Precipitation in Winter (left panel) and Summer (right panel), 2025.}
	\label{fig:Contour_Plots}
\end{figure*}

Figure~\ref{fig:Contour_Plots} displays state-wise bivariate kernel-density contours of daily temperature (x-axis) and precipitation (y-axis) for the West and East South Central U.S. (AL, AR, LA, MS, OK, TN, TX). Winter spans 1 December 2024-28 February 2025, and summer spans 1 June-31 August 2025. Within each panel, contours represent high-density regions of the joint distribution, while the marginal curves summarize the corresponding univariate temperature and precipitation distributions.

From winter to summer, the bivariate temperature-precipitation density contours shift substantially rightward, and the state-specific centers that span a wide winter temperature range collapse into a narrow, uniformly hot summer band, indicating that temperature provides stronger between-state separation in winter than in summer. Precipitation exhibits a contrasting seasonal behavior; while densities remain sharply concentrated near zero in both seasons, summer shows a pronounced heavy right tail, with more frequent and larger high-precipitation days, whereas winter wet-day regions are smaller and contribute less to the overall mass. The contour geometry further suggests a change in co-variability, with winter contours elongated along the temperature axis (larger day-to-day temperature variability at near-zero precipitation) and summer contours more vertically diffuse, particularly for Gulf-influenced states, reflecting increased rainfall variability at similarly hot temperatures. Consistent with this, Gulf states (LA, MS, AL, and southeastern TX) occupy the warmer-and-wetter winter regime and display the strongest summer precipitation tails, while interior and Southern Plains states (OK and inland TX) are cooler and drier in winter and retain comparatively thin summer precipitation tails despite converging to the common hot temperature mode. In general, the contours and marginals jointly summarize how the typical conditions and extremes of each state differ by season and location.


Table~\ref{tab:Most_influential_variable_selection} decomposes the significant multivariate state–state differences into their temperature and precipitation components. For each significant pair, we evaluate daily mean temperature and precipitation separately and report the corresponding BH‑adjusted \textit{p}-values. This assessment clarifies whether the observed differences are driven by temperature, precipitation, or a combination of both.

\begin{table*}[!ht]
\centering
\caption{Variable contributions to pairwise state differences.}
\begin{tabular}{lcccc}
\toprule
Season & Pair & $p_{\text{BH}}$(TAVG) & $p_{\text{BH}}$(PRCP) & Driver \\
\midrule
    
     & AL vs TN & 0.005 & 0.028 & Both \\
    & MS vs TN & $<0.001$ & 0.005 & Both \\
  Winter  & MS vs TX & 0.046 & 0.005 & Both \\
    & OK vs TN & 0.006 & 0.147 & Temperature \\
    & TN vs TX & $<0.001$ & 0.111 & Temperature \\
    \hline
     & AL vs LA & 0.049 & 0.069 & Temperature\\
    & AL vs MS & 0.015 & 0.257 & Temperature \\
    & AL vs OK & 0.018 & 0.030 & Both \\
    & AL vs TX & 0.004 & 0.069 & Temperature \\
    & LA vs MS & 0.036 & 0.257 & Temperature \\
   Summer & LA vs OK & 0.049 & 0.098 & Temperature\\
    & LA vs TN & 0.049 & 0.069 & Temperature\\
    & MS vs OK & $<0.001$ & 0.011 & Both \\
    & MS vs TN & 0.008 & 0.098 & Temperature \\
    & OK vs TX & 0.011 & 0.029 & Both \\
    & TN vs TX & $<0.001$ & 0.029 & Both\\
\bottomrule
\end{tabular}
\label{tab:Most_influential_variable_selection}
\end{table*}

\newpage
\clearpage
\section{Gaussian Mixture Model - Spatiotemporal Dynamics of Climate Clusters}\label{secA1}

\begin{figure*}[!ht]
    \centering
    \includegraphics[width=1.0\linewidth]{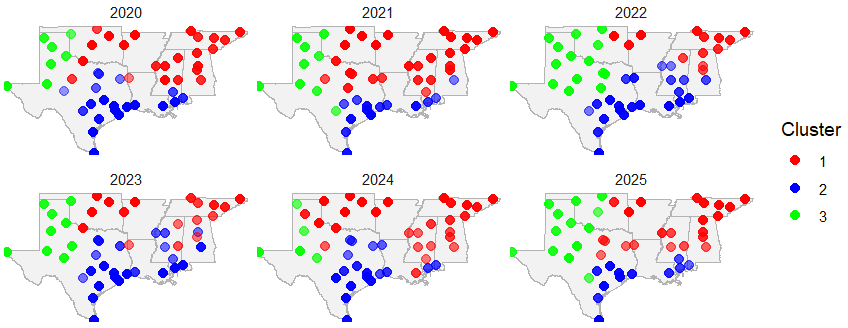}
    \caption{Gaussian Mixture Model clustering of station-winter climate features.}
    \label{fig:winter_clustering2020_2025}
\end{figure*}

Figure~\ref{fig:winter_clustering2020_2025} presents the GMM clustering results for winter in the study region over the years 2020–2025.

\begin{figure*}[!ht]
    \centering
    \includegraphics[width = 1.0\linewidth]{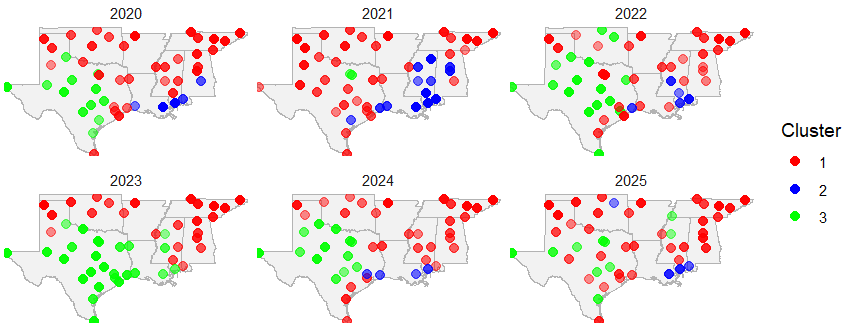}
    \caption{Gaussian Mixture Model clustering of station-summer climate features.}
    \label{fig:summer_clustering2020_2025}
\end{figure*}

Figure~\ref{fig:summer_clustering2020_2025} illustrates the summer GMM clusters identified across the study region for the period 2020–2025.

\begin{figure*}[!ht]
     \centering
     \begin{subfigure}[b]{0.49\textwidth}
         \centering
       \includegraphics[width=1.0\textwidth]{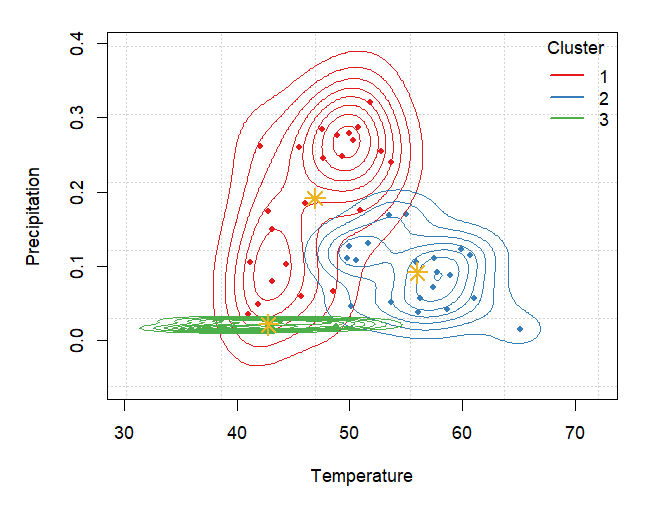}
       \caption{Winter 2020}
     \end{subfigure}  
     \begin{subfigure}[b]{0.49\textwidth}
         \centering
       \includegraphics[width=1.0\textwidth]{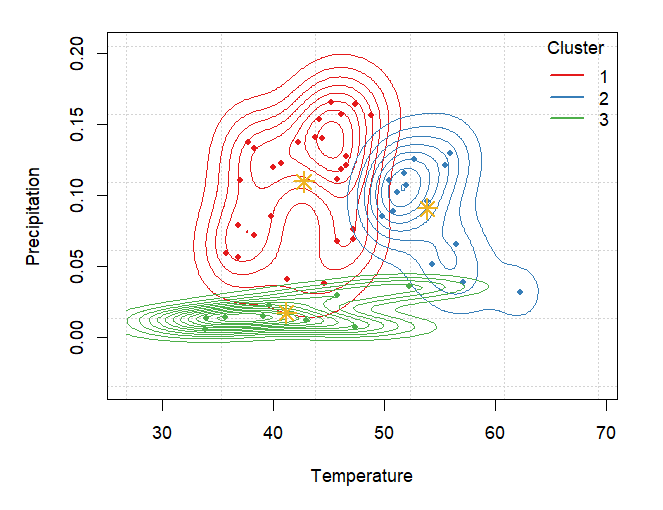}
       \caption{Winter 2021.}
     \end{subfigure}
          \begin{subfigure}[b]{0.49\textwidth}
         \centering
       \includegraphics[width=1.0\textwidth]{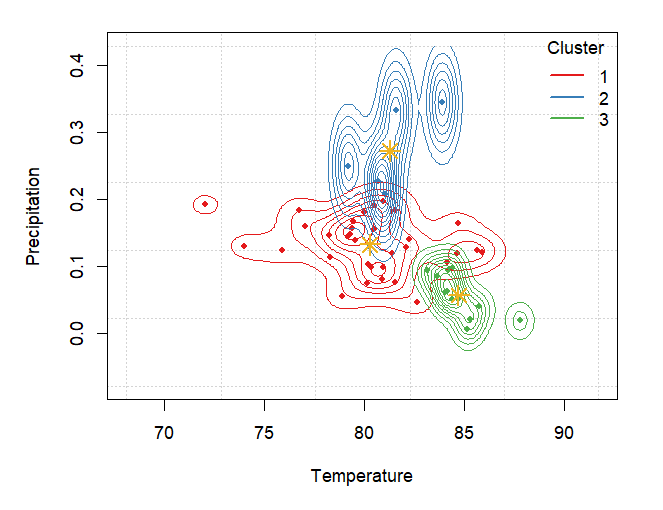}
       \caption{Summer 2020}
     \end{subfigure}
       \begin{subfigure}[b]{0.49\textwidth}
         \centering
       \includegraphics[width=1.0\textwidth]{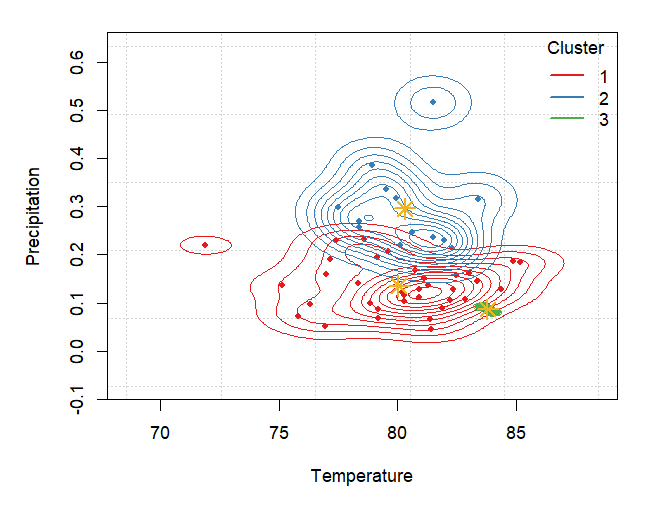}
       \caption{Summer 2021.}
     \end{subfigure}
\caption {Distribution of climate variables in different clusters}
\label{fig:smoothfun}
\end{figure*}

\begin{figure*}[!ht]
     \centering
     \begin{subfigure}[b]{0.49\textwidth}
         \centering
       \includegraphics[width=1.0\textwidth]{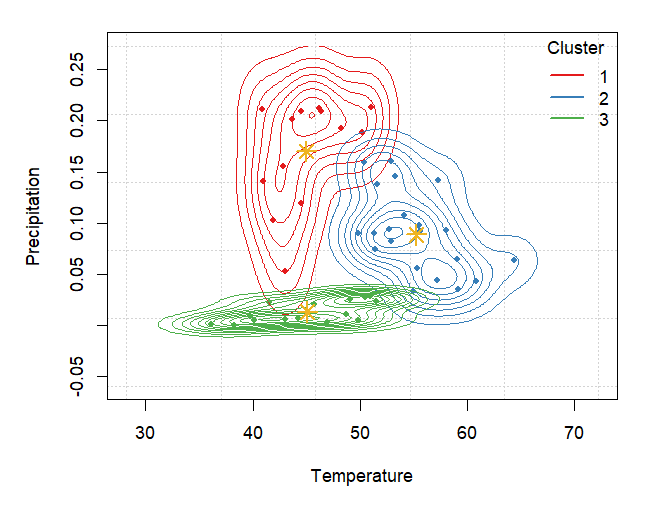}
       \caption{Winter 2022}
     \end{subfigure}  
     \begin{subfigure}[b]{0.49\textwidth}
         \centering
       \includegraphics[width=1.0\textwidth]{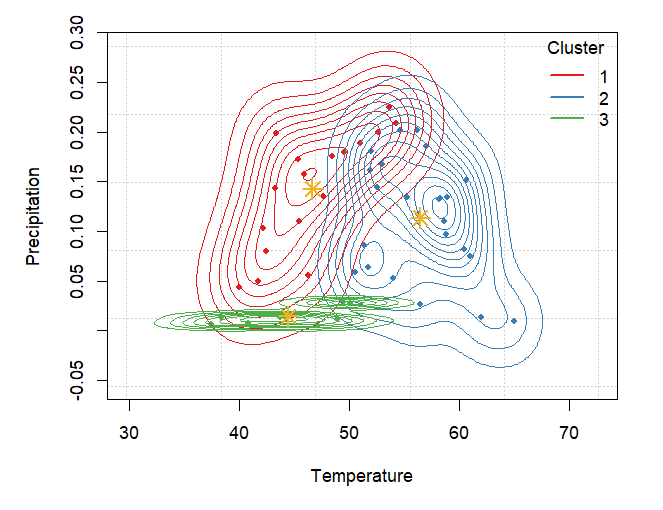}
       \caption{Winter 2023.}
     \end{subfigure}
          \begin{subfigure}[b]{0.49\textwidth}
         \centering
       \includegraphics[width=1.0\textwidth]{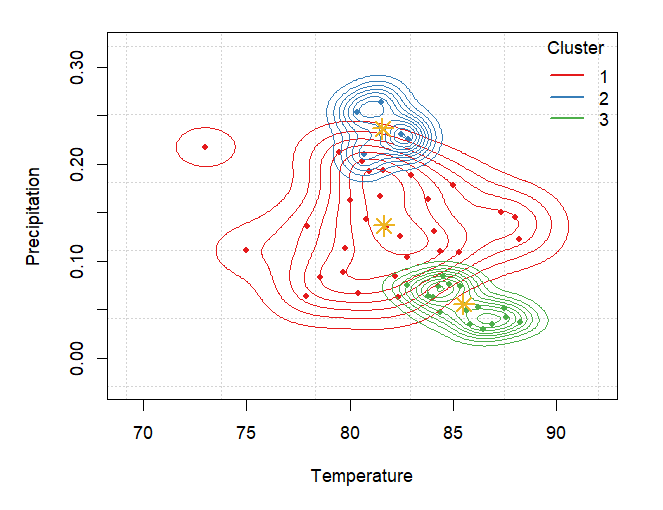}
       \caption{Summer 2022}
     \end{subfigure}
       \begin{subfigure}[b]{0.49\textwidth}
         \centering
       \includegraphics[width=1.0\textwidth]{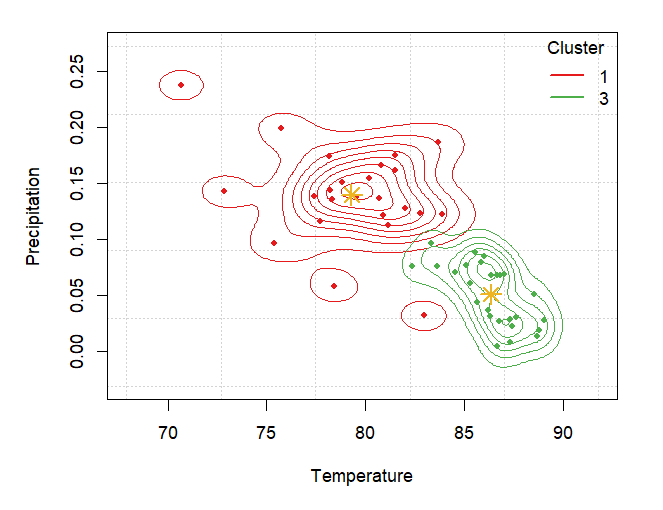}
       \caption{Summer 2023.}
     \end{subfigure}
\caption {Distribution of climate variables in different clusters}
\label{fig:cluster11}
\end{figure*}

\begin{figure*}[!ht]
     \centering
     \begin{subfigure}[b]{0.49\textwidth}
         \centering
       \includegraphics[width=1.0\textwidth]{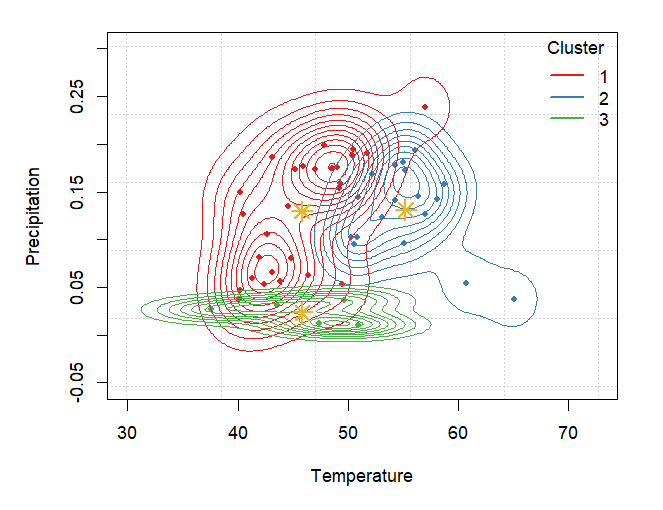}
       \caption{Winter 2024}
     \end{subfigure}  
          \begin{subfigure}[b]{0.49\textwidth}
         \centering
       \includegraphics[width=1.0\textwidth]{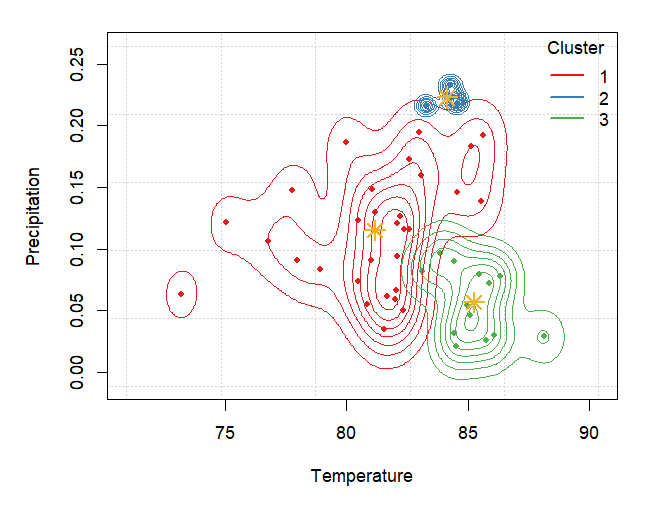}
       \caption{Summer 2024.}
     \end{subfigure}
\caption {Distribution of climate variables in different clusters}
\label{fig:cluster12} 
\end{figure*}

Figures~\ref{fig:smoothfun}-~\ref{fig:cluster12} summarize the joint temperature-precipitation structure of the GMM clusters using bivariate kernel-density contours for winter and summer in 2020-2024. In winters (2020, 2021, 2022, 2023, and 2024), clusters are distinguished mainly by temperature. Cluster 3 concentrates on the coolest conditions with precipitation near zero; Cluster 1 spans cool-to-mild temperatures and exhibits the widest precipitation variability (capturing the wetter winter episodes); and Cluster 2 shifts to the warmest winter regime with moderate precipitation spread. In summers, temperatures collapse into a narrow hot band, so the clustering is driven primarily by the vertical extent of precipitation. Cluster 3 remains the dry-hot regime with rainfall concentrated near zero, Cluster 1 represents typical hot days with moderate rainfall variability, and Cluster 2 isolates the wettest conditions with the heaviest high-precipitation tail. Year-to-year variation is expressed primarily through changes in the relative dominance, spread, and tail thickness of the clusters, rather than changes in the basic seasonal organization. In particular, the wet-regime structure is most variable over years, with differences in how distinctly the high-precipitation cluster separates and how strongly it extends into the upper tail. Overall, these figures indicate stable winter temperature-driven partitioning and more precipitation-driven summer partitioning, with interannual variability arising primarily from precipitation variability and extremes.


\end{appendices}


\newpage
\clearpage

\bibliographystyle{apalike}
\bibliography{climate}

@article{Gorecki10092017,
author = {T. G{\'o}recki and {\L}. Smaga},
title = {Multivariate analysis of variance for functional data},
journal = {Journal of Applied Statistics},
volume = {44},
number = {12},
pages = {2172--2189},
year = {2017},
publisher = {Taylor \& Francis},
doi = {10.1080/02664763.2016.1247791}
}

@article{shmueli2010explain,
  title={To explain or to predict?},
  author={Shmueli, Galit},
  journal={Statistical science},
  pages={289--310},
  year={2010},
  publisher={JSTOR}
}

@article{lesk2022_compound_heat_moisture,
  author  = {Lesk, Carolyn and Anderson, William and Rigden, Amanda and others},
  title   = {Compound heat and moisture extreme impacts on global crop yields under climate change},
  journal = {Nature Reviews Earth \& Environment},
  volume  = {3},
  pages   = {872--889},
  year    = {2022},
  doi     = {10.1038/s43017-022-00368-8}
}

@article{Megan_2020,
author = {Megan C. Kirchmeier-Young  and Xuebin Zhang },
title = {Human influence has intensified extreme precipitation in North America},
journal = {Proceedings of the National Academy of Sciences},
volume = {117},
number = {24},
pages = {13308-13313},
year = {2020},
doi = {10.1073/pnas.1921628117}}

@article {Martin_2021,
      author = "Martin Hoerling and Lesley Smith and Xiao-Wei Quan and Jon Eischeid and Joseph Barsugli and Henry F. Diaz",
      title = "Explaining the Spatial Pattern of U.S. Extreme Daily Precipitation Change",
      journal = "Journal of Climate",
      year = "2021",
      publisher = "American Meteorological Society",
      address = "Boston MA, USA",
      volume = "34",
      number = "7",
      doi = "10.1175/JCLI-D-20-0666.1",
      pages=      "2759 - 2775"
}

@article{Danielle_2022,
author = {Danielle Touma  and Samantha Stevenson  and Daniel L. Swain  and Deepti Singh  and Dmitri A. Kalashnikov  and Xingying Huang },
title = {Climate change increases risk of extreme rainfall following wildfire in the western United States},
journal = {Science Advances},
volume = {8},
number = {13},
pages = {eabm0320},
year = {2022},
doi = {10.1126/sciadv.abm0320}}

@article{gori2022_tc_rainfall_surge,
  author  = {Gori, Andrea and Lin, Ning and Xi, Dan and others},
  title   = {Tropical cyclone climatology change greatly exacerbates {US} extreme rainfall--surge hazard},
  journal = {Nature Climate Change},
  volume  = {12},
  pages   = {171--178},
  year    = {2022},
  doi     = {10.1038/s41558-021-01272-7}
}

@article{knutti2016climate,
  author  = {Knutti, Reto and Rugenstein, Maria A. A. and Hegerl, Gabriele C.},
  title   = {Beyond equilibrium climate sensitivity},
  journal = {Nature Geoscience},
  year    = {2016},
  volume  = {10},
  pages   = {727--736}
}

@article{Tania_2020,
author = {Lopez-Cantu, Tania and Prein, Andreas F. and Samaras, Constantine},
title = {Uncertainties in Future U.S. Extreme Precipitation From Downscaled Climate Projections},
journal = {Geophysical Research Letters},
volume = {47},
number = {9},
pages = {e2019GL086797},
doi = {https://doi.org/10.1029/2019GL086797},
year = {2020}
}

@article{Dey02012016,
author = {Asim Kumer Dey and Kumer Pial Das},
title = {Modeling Extreme Hurricane Damage Using the Generalized Pareto Distribution},
journal = {American Journal of Mathematical and Management Sciences},
volume = {35},
number = {1},
pages = {55--66},
year = {2016},
publisher = {Taylor \& Francis},
doi = {10.1080/01966324.2015.1075926}

}

@article{Miniussi_2020,
author = {Miniussi, Arianna and Villarini, Gabriele and Marani, Marco},
title = {Analyses Through the Metastatistical Extreme Value Distribution Identify Contributions of Tropical Cyclones to Rainfall Extremes in the Eastern United States},
journal = {Geophysical Research Letters},
volume = {47},
number = {7},
pages = {e2020GL087238},
doi = {https://doi.org/10.1029/2020GL087238},
year = {2020}
}

@article{franzke2015climate,
  author  = {Franzke, Christian L. E. and O'Kane, Terence J. and Berner, Judith and Williams, Paul D. and Lucarini, Valerio},
  title   = {Stochastic climate theory and modeling},
  journal = {Wiley Interdisciplinary Reviews: Climate Change},
  year    = {2015},
  volume  = {6},
  pages   = {63--78}
}

@article{ECK2020108053,
title = {Influence of growing season temperature and precipitation anomalies on crop yield in the southeastern United States},
journal = {Agricultural and Forest Meteorology},
volume = {291},
pages = {108053},
year = {2020},
issn = {0168-1923},
doi = {https://doi.org/10.1016/j.agrformet.2020.108053},
url = {https://www.sciencedirect.com/science/article/pii/S0168192320301556},
author = {Montana A. Eck and Andrew R. Murray and Ashley R. Ward and Charles E. Konrad}
}

@article{Cardell_2020,
author = {Cardell, Maria F. and Amengual, Arnau and Romero, Romualdo and Ramis, Climent},
title = {Future extremes of temperature and precipitation in Europe derived from a combination of dynamical and statistical approaches},
journal = {International Journal of Climatology},
volume = {40},
number = {11},
pages = {4800-4827},
doi = {https://doi.org/10.1002/joc.6490},
year = {2020}
}

@article{Zhong02012025,
author = {Peng Zhong and Manuela Brunner and Thomas Opitz and Raphaël Huser},
title = {Spatial Modeling and Future Projection of Extreme Precipitation Extents},
journal = {Journal of the American Statistical Association},
volume = {120},
number = {549},
pages = {80--95},
year = {2025},
publisher = {Taylor \& Francis},
doi = {10.1080/01621459.2024.2408045}

}

@article{Gorecki_2019,
author = {G{\'o}recki, Tomasz and Smaga, {\L}ukasz},
year = {2019},
month = {06},
pages = {},
title = {fdANOVA: an R software package for analysis of variance for univariate and multivariate functional data},
volume = {34},
journal = {Computational Statistics},
doi = {10.1007/s00180-018-0842-7}
}

@article{Dempster_2018,
    author = {Dempster, A. P. and Laird, N. M. and Rubin, D. B.},
    title = {Maximum Likelihood from Incomplete Data Via the {EM} Algorithm},
    journal = {Journal of the Royal Statistical Society: Series B (Methodological)},
    volume = {39},
    number = {1},
    pages = {1-22},
    year = {2018},
    month = {12},
    issn = {0035-9246},
    doi = {10.1111/j.2517-6161.1977.tb01600.x}
}

@article{Wilks1932,
  author = {Wilks, S. S.},
  title  = {Certain Generalizations in the Analysis of Variance},
  journal= {Biometrika},
  year   = {1932},
  volume = {24},
  number = {3-4},
  pages  = {471--494},
  doi    = {10.1093/biomet/24.3-4.471}
}

@article{Pillai1955,
  author = {Pillai, K. C. S.},
  title  = {Some New Test Criteria in Multivariate Analysis},
  journal= {The Annals of Mathematical Statistics},
  year   = {1955},
  volume = {26},
  number = {1},
  pages  = {117--121},
  doi    = {10.1214/aoms/1177728599}
}

@book{Anderson1958,
  author = {Anderson, T. W.},
  title  = {An Introduction to Multivariate Statistical Analysis},
  year   = {1958},
  publisher = {Wiley}
}

@article{TODOROV201037,
title = {Robust statistic for the one-way MANOVA},
journal = {Computational Statistics \& Data Analysis},
volume = {54},
number = {1},
pages = {37-48},
year = {2010},
issn = {0167-9473},
doi = {https://doi.org/10.1016/j.csda.2009.08.015},
url = {https://www.sciencedirect.com/science/article/pii/S0167947309003090},
author = {Valentin Todorov and Peter Filzmoser}
}

@article{benjamini1995controlling,
  title={Controlling the false discovery rate: a practical and powerful approach to multiple testing},
  author={Benjamini, Yoav and Hochberg, Yosef},
  journal={Journal of the Royal statistical society: series B (Methodological)},
  volume={57},
  number={1},
  pages={289--300},
  year={1995},
  publisher={Wiley Online Library}
}

@misc{NOAA_NCEI_CDO_2025,
  author       = {{NCEI}},
  title        = {Climate Data Online (CDO)},
  year         = {2025},
  note         = {Accessed 2025-11-07},
  howpublished = {\url{https://www.ncei.noaa.gov/cdo-web/}}
}

@book{schumaker2007spline,
  title     = {Spline Functions: Basic Theory},
  author    = {Schumaker, Larry L.},
  publisher = {Cambridge University Press},
  year      = {2007},
  edition   = {3rd}
}

@book{deboor1978practical,
  title     = {A Practical Guide to Splines},
  author    = {de Boor, Carl},
  year      = {1978},
  publisher = {Springer},
  address   = {New York},
  series    = {Applied Mathematical Sciences},
  volume    = {27}
}

@article{Fraley01062002,
author = {Chris Fraley and Adrian E Raftery},
title = {Model-Based Clustering, Discriminant Analysis, and Density Estimation},
journal = {Journal of the American Statistical Association},
volume = {97},
number = {458},
pages = {611--631},
year = {2002},
publisher = {ASA Website},
doi = {10.1198/016214502760047131},
URL = {     https://doi.org/10.1198/016214502760047131
},
eprint = {https://doi.org/10.1198/016214502760047131
}}

@Book{mclust,
    title = {Model-Based Clustering, Classification, and Density
      Estimation Using {mclust} in {R}},
    author = {Luca Scrucca and Chris Fraley and T. Brendan Murphy and
      Adrian E. Raftery},
    publisher = {Chapman and Hall/CRC},
    isbn = {978-1032234953},
    doi = {10.1201/9781003277965},
    year = {2023},
    url = {https://mclust-org.github.io/book/},
  }

@article{Meng_2002,
    author = {Meng, Xiao-Li and Van Dyk, David},
    title = {The EM Algorithm—an Old Folk-song Sung to a Fast New Tune},
    journal = {Journal of the Royal Statistical Society: Series B (Methodological)},
    volume = {59},
    number = {3},
    pages = {511-567},
    year = {2002},
    month = {01},
    issn = {0035-9246},
    doi = {10.1111/1467-9868.00082}
}

@book{bishop2006prml,
  author    = {Bishop, Christopher M.},
  title     = {Pattern Recognition and Machine Learning},
  publisher = {Springer},
  address   = {New York},
  year      = {2006},
  series    = {Information Science and Statistics}
}

@article{region2017quarterly,
  title={Quarterly Climate Impacts and Outlook},
  author={Region, Great Lakes},
  journal={Dep},
  year={2017}
}

@incollection{ipcc_ar6_wgi_spm_2021,
  author    = {{IPCC}},
  title     = {Summary for Policymakers},
  booktitle = {Climate Change 2021: The Physical Science Basis. Contribution of Working Group I to the Sixth Assessment Report of the Intergovernmental Panel on Climate Change},
  editor    = {Masson-Delmotte, V. and Zhai, P. and Pirani, A. and Connors, S. L. and P{\'e}an, C. and Berger, S. and Caud, N. and Chen, Y. and Goldfarb, L. and Gomis, M. I. and Huang, M. and Leitzell, K. and Lonnoy, E. and Matthews, J. B. R. and Maycock, T. K. and Waterfield, T. and Yelek{\c{c}}i, O. and Yu, R. and Zhou, B.},
  publisher = {Cambridge University Press},
  year      = {2021},
  pages     = {3--32},
  doi       = {10.1017/9781009157896.001}
}

@incollection{mcpherson2023_nca5_ch26,
  author    = {McPherson, Renee A. and Fay, Philip A. and Alvarez, Susan G. and Bertrand, Darrian and Broadbent, Taylor L. and Bruno, Tianna and Fares, Ali and McCullough, Brian and Moore, Georgianne W. and Moorhead, Bee and Pati{\~n}o, Laura and Petersen, Alexander and Smith, Nicholas G. and Steiner, Jean L. and Taylor, April and Warziniack, Travis},
  title     = {Ch. 26. Southern Great Plains},
  booktitle = {Fifth National Climate Assessment},
  editor    = {Crimmins, A. R. and Avery, C. W. and Easterling, D. R. and Kunkel, K. E. and Stewart, B. C. and Maycock, T. K.},
  publisher = {U.S. Global Change Research Program},
  address   = {Washington, DC, USA},
  year      = {2023},
  doi       = {10.7930/NCA5.2023.CH26}
}

@article{gil2022temperature,
  title={Temperature and precipitation in the US states: long memory, persistence, and time trend},
  author={Gil-Alana, Luis A and Gupta, Rangan and Sauci, Laura and Carmona-Gonz{\'a}lez, Nieves},
  journal={Theoretical and Applied Climatology},
  volume={150},
  number={3},
  pages={1731--1744},
  year={2022},
  publisher={Springer}
}

@article{lai2020use,
  title={Use of the autoregressive integrated moving average (ARIMA) model to forecast near-term regional temperature and precipitation},
  author={Lai, Yuchuan and Dzombak, David A},
  journal={Weather and forecasting},
  volume={35},
  number={3},
  pages={959--976},
  year={2020}
}

@article{armal2018trends,
  title={Trends in extreme rainfall frequency in the contiguous United States: Attribution to climate change and climate variability modes},
  author={Armal, Saman and Devineni, Naresh and Khanbilvardi, Reza},
  journal={Journal of Climate},
  volume={31},
  number={1},
  pages={369--385},
  year={2018}
}

@article{cooley2010spatial,
  title={Spatial hierarchical modeling of precipitation extremes from a regional climate model},
  author={Cooley, Daniel and Sain, Stephan R},
  journal={Journal of agricultural, biological, and environmental statistics},
  volume={15},
  number={3},
  pages={381--402},
  year={2010},
  publisher={Springer}
}

@article{fard2023spatio,
  title={Spatio-temporal interpolation and delineation of extreme heat events in California between 2017 and 2021},
  author={Fard, Pedram and Chung, Ming Kei Jake and Estiri, Hossein and Patel, Chirag J},
  journal={Environmental research},
  volume={237},
  pages={116984},
  year={2023},
  publisher={Elsevier}
}

@article{bunting2024heat,
  title={What is a heat wave: A survey and literature synthesis of heat wave definitions across the United States},
  author={Bunting, Erin L and Tolmanov, Vasily and Keellings, David},
  journal={PLOS Climate},
  volume={3},
  number={9},
  pages={e0000468},
  year={2024},
  publisher={Public Library of Science San Francisco, CA USA}
}

@article{DHARMARATHNE2024102123,
title = {Adapting cities to the surge: A comprehensive review of climate-induced urban flooding},
journal = {Results in Engineering},
volume = {22},
pages = {102123},
year = {2024},
issn = {2590-1230},
doi = {https://doi.org/10.1016/j.rineng.2024.102123},
url = {https://www.sciencedirect.com/science/article/pii/S2590123024003773},
author = {Gangani Dharmarathne and A.O. Waduge and Madhusha Bogahawaththa and Upaka Rathnayake and D.P.P. Meddage}
}

@article{naz2019effects,
  title={Effects of Climate Change on Human Behaviour: A People Perspective},
  author={Naz, Rahat and Shah, Mussawar and Jamal, Humera and Khan, Younas},
  journal={J. Appl. Environ. Biol. Sci},
  volume={9},
  number={5},
  pages={1--10},
  year={2019}
}

@Article{cli4010010,
AUTHOR = {Chattopadhyay, Somsubhra and Edwards, Dwayne R.},
TITLE = {Long-Term Trend Analysis of Precipitation and Air Temperature for Kentucky, United States},
JOURNAL = {Climate},
VOLUME = {4},
YEAR = {2016},
NUMBER = {1},
ARTICLE-NUMBER = {10},
URL = {https://www.mdpi.com/2225-1154/4/1/10},
ISSN = {2225-1154},
DOI = {10.3390/cli4010010}
}

@article{shen2016six,
  title={Six temperature and precipitation regimes of the contiguous United States between 1895 and 2010: a statistical inference study},
  author={Shen, Samuel SP and Wied, Olaf and Weithmann, Alexander and Regele, Tobias and Bailey, Barbara A and Lawrimore, Jay H},
  journal={Theoretical and Applied Climatology},
  volume={125},
  number={1},
  pages={197--211},
  year={2016},
  publisher={Springer}
}

@article{barriopedro2023heat,
  title={Heat waves: Physical understanding and scientific challenges},
  author={Barriopedro, David and Garc{\'\i}a-Herrera, R and Ord{\'o}nez, Carlos and Miralles, Diego G and Salcedo-Sanz, Sancho},
  journal={Reviews of Geophysics},
  volume={61},
  number={2},
  pages={e2022RG000780},
  year={2023},
  publisher={Wiley Online Library}
}

@article{perkins2013measurement,
  title={On the measurement of heat waves},
  author={Perkins, Sarah E and Alexander, Lisa V},
  journal={Journal of climate},
  volume={26},
  number={13},
  pages={4500--4517},
  year={2013}
}

\end{document}